\documentclass[twocolumn,preprintnumbers,amsmath,amssymb,superscriptaddress]{revtex4}
\usepackage{graphicx}
\usepackage{dcolumn}
\usepackage{color}
\usepackage{bm}
\usepackage{amsmath}

\begin{document}

\title{Quantum Enhanced Phase Retrieval}

\author{Liat Liberman\footnote{These authors contributed equally to this work}}
\author{Yonatan Israel$^{\ast}$} 
\affiliation{Department of Physics of Complex Systems, Weizmann Institute of Science, Rehovot 76100, Israel}

\author{Eilon Poem}
\affiliation{Clarendon Laboratory, University of Oxford, Parks Road, Oxford OX1 3PU, UK}

\author{Yaron Silberberg}
\affiliation{Department of Physics of Complex Systems, Weizmann Institute of Science, Rehovot 76100, Israel}





\begin{abstract}
The retrieval of phases from intensity measurements is a key process in many fields in science, from optical microscopy to x-ray crystallography. Here we study phase retrieval of a one-dimensional multi-phase object that is illuminated by quantum states of light. We generalize the iterative Gerchberg-Saxton algorithm to photon correlation measurements on the output plane, rather than the standard intensity measurements. We report a numerical comparison of classical and quantum phase retrieval of a small one-dimensional object of discrete phases from its far-field diffraction. While the classical algorithm was ambiguous and often converged to wrong solutions, quantum light produced a unique reconstruction with smaller errors and faster convergence. We attribute these improvements to a larger Hilbert space that constrains the algorithm.
\end{abstract}
\maketitle



\section{Introduction}

Quantum states of light have been widely explored in recent years for their ability to offer considerable enhancement in measurement sensitivity over classical ones \cite{QuantumMetrologyReview2011}. Quantum states were mostly considered for enhancing the sensitivity of the measurement of a single optical phase using an optical interferometer. Recently, the problem of simultaneous estimation of several optical phases using quantum light was investigated \cite{spagnolo2012,MultiPhaseWalmsleyPRL2013}. Here we investigate an iterative phase-retrieval technique for the estimation of one-dimensional phase objects with quantum states of light, and show that the technique is more robust than its classical version.

The problem of phase retrieval is one of great scientific interest, which arises when the intensity recorded in the far-field is used to determine the phase structure of an object, information which is otherwise undetected. It stems from the fact that detectors record the intensity of waves, while often the important information is encoded in their phases. Phase retrieval has been intensively investigated \cite{shechtmanPR} and found applications across many fields of science, from astronomy (wave-front sensing) \cite{PR_astronomy} to nanotechnology (x-ray crystallography and electron microscopy) \cite{MiaoXray,ElectronMicroscopyBook}. In the most common scenario, the far-field diffraction intensity pattern is measured; This measured far-field intensity, together with known constrains on the illumination (e.g. its intensity profile) are used in an iterative algorithm to derive the phase structure of the object \cite{GerchbergAndSaxton1972}. It is known, however, that phase retrieval has certain limitation; for example, phase-retrieval of one-dimensional objects is problematic, and often leads to multiple ambiguous solutions \cite{shechtmanPR}.


\begin{figure}[b!]
	\centering
	\includegraphics[width=.85\columnwidth]{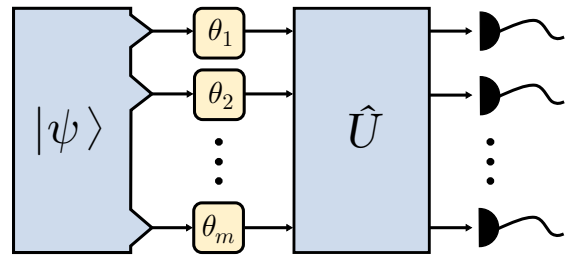}
    \caption{A schematic description of phase retrieval using quantum states. A quantum state $|\psi\rangle$, entangled over $m$ modes, is input to a multi-mode interferometric system. The state passes through $m$ phases denoted by $\vec{\theta} = \{\theta_1,\theta_2,\dots,\theta_m\}$, and followed by a transformation $\hat{U}$. Photon correlation measurement is carried out on the output state. \label{fig:system}}
\end{figure}

Here we study the use of quantum states of light to measure a one-dimensional multi-phase object, as shown schematically in Fig. \ref{fig:system}, using a phase-retrieval approach. We present a protocol that utilizes entangled states to reconstruct multiple phases simultaneously using an iterative error-reduction algorithm. It is shown by our numerical results that a quantum approach has a few advantages over the approaches that use classical light. First, phase retrieval using quantum light can be unambiguous, as it reaches the single, correct solution. The algorithm also converges much faster than in the classical case. Furthermore, the use of quantum states can enhance the sensitivity of the retrieved phases over cases where classical light is used, for the same number of photons probing the system. These quantum enhanced capabilities are already revealed for two-photon entangled states, and are particularly important when probing delicate samples which are sensitive to illumination intensity, such as biological samples \cite{BiologicalMicroscopyReview2003}, quantum gases \cite{QNDofQuantumGasesNatPhys2007}, and atomic ensembles \cite{EntandledProbingDelicateMaterialNatPhoton2012}. With recent advances in generating quantum states in x-ray \cite{XRAY-PDC_ShwartzPRL2012}, their application for phase retrieval holds great promise.

An important milestone in the field of phase retrieval was the iterative algorithm of Gerchberg and Saxton \cite{GerchbergAndSaxton1972}, that reduces the error in the phase object with every iteration. It is doing so by iterating between the input plane $E_{in}(x)$, and the output plane $\tilde{E}_{out}(u)$, related by the known transformation of the system, and applying the information obtained from the intensity of the input plane and the output intensity measurement at each iteration of the algorithm.
In the common case of far-field diffraction, the transformation between the input and output planes is the Fourier transform, where $x$ and $u$ are the coordinates along the input and output planes, respectively. Similarly, phase retrieval can be applied to the frequency-time domains, where the temporal field $E_{in}(t)$ is related to its spectral one $\tilde{E}_{out}(\omega)$ by a Fourier transformation, and the problem relates to determining temporal amplitudes and phases from spectral power measurements \cite{trebino93}.
However, phase reconstruction is not unique in general. As the amplitudes of the input and output planes $|E_{in}(x)|$, and $|\tilde{E}_{out}(u)|$ are restricted by their measured intensities, there might be additional solutions for the phase image which are incorrect. Some examples include shifted images $E_{in}(x-x_0)$, mirror images $E_{in}^*(-x)$, and global phases $e^{i\Phi}E_{in}(x)$, which are exact solutions of the phase retrieval problem, yet there may exist many other non-trivial exact solutions as well \cite{Walther,sanz1985}, which to the best of our knowledge were never analyzed.

Quantum enhanced phase retrieval uses quantum states to probe the object, as we outline in the next section. In the far-field, instead of intensity measurements, we employ measurement of photon correlations on the output quantum states. Using these measurements and the knowledge of the input state, we describe the algorithm used for the retrieval of the phases of the object. We then describe a specific example, where we also discuss in some detail the sensitivity of using quantum light for phase retrieval.

\section{Methodology}
\label{sec:methodology}

Let us consider the problem of estimating a one-dimensional object of multiple phases, probed by a quantum state of light, as shown in Fig. \ref{fig:system}.
A phase object is characterized by a set of $m$ unknown phases, $\vec{\theta}=\{\theta_1,\dots,\theta_m\}$.

\subsection{Quantum light}

An initial pure state of $N$ photons in $m$ modes has the form
\begin{equation}
    \begin{aligned} \label{eq:psi_input}
       |\psi\rangle = \sum_{k=1}^D \alpha_k |n_1^{(k)},n_2^{(k)},\dots,n_m^{(k)}\rangle = \sum_{k=1}^D \alpha_k |\vec{n}^{(k)}\rangle,
	\end{aligned}
\end{equation}
where $\vec{n}^{(k)}$ is a vector of length $m$ with photon number components $n_{x}^{(k)}$ in each mode $x$ and for each configuration $k$, such that $\sum_{x=1}^m n_x^{(k)} = N$. The set of amplitudes $\vec{\alpha} = \{\alpha_k\}$ (where $k=1,\dots,D$) is normalized $\sum_{k=1}^D |\alpha_k|^2 =1$, and the total number of configurations is $D = \binom{N+m-1}{N}$.

After passing through the phase object the state accrues $m$ phases, as described by the unitary transformation $\hat{U}_{\vec{\theta}} = \exp(\imath \sum_{x=1}^m \theta_x \hat{n}_x)$, where $\hat{n}_x$ is the number operator for mode $x$. The state in Eq. \ref{eq:psi_input} then becomes
\begin{equation}
\label{eq:psi_theta}
|\psi_{\vec{\theta}}\rangle = \hat{U}_{\vec{\theta}} |\psi\rangle = \sum_{k=1}^D \alpha_k e^{\imath \phi_k} |\vec{n}^{(k)}\rangle,
\end{equation}
where the set of phases accrued by the state, $\vec{\phi} = \{\phi_k\}$ is related to the object phases $\vec{\theta}$ by $\phi_k = \vec{\theta} \cdot \vec{n}^{(k)}$.

Next, this quantum state undergoes a transformation, most commonly Fourier transformation via diffraction, which transforms the state in Eq. \ref{eq:psi_theta} to the final state at the output
\begin{equation}\label{eq:psi_F}
    |\psi_{F}\rangle = \hat{U} |\psi_{\vec{\theta}}\rangle  = \sum_{t=1}^D \beta_t |\vec{n}^{(t)}\rangle.
\end{equation}
In Eq. \ref{eq:psi_F} we assumed that the transformation described by $\hat{U}$ is unitary; it can therefore be represented by a unitary $m\times m$ matrix $U$. Using such operation transforms the photon creation operators, in any mode $x$, by $\hat{a}^{\dagger}_x \rightarrow \sum_{x=1}^m [U]_{x,y} \hat{a}^{\dagger}_y$. One example of such an operation is the discrete Fourier transform (DFT), which we use as the transformation to the far-field plane. The DFT is represented by an $m \times m$ matrix $[U]_{x,y}=\exp(\imath 2 \pi (x-1)(y-1)/m)/\sqrt{m}$ \cite{marek_MultiportBS_1997}. The set of amplitudes denoted by $\vec{\beta}$ in Eq. \ref{eq:psi_F} can be calculated as a function of the input state amplitudes $\vec{\alpha}$ of Eq. \ref{eq:psi_input},
\begin{equation}\label{eq:psi_perm}
	\beta_t = \langle \vec{n}^{(t)}|\hat{U}|\psi_{\vec{\theta}} \rangle = \sum_{k=1}^D \frac{\alpha_k e^{\imath \phi_k} \textrm{Per}(V_{k,t})} {\sqrt{\prod_{x=1}^{m}(n_{x}^{(k)})!\prod_{y=1}^{m}(n_{y}^{(t)})!}},
\end{equation}
where $V_{k,t}$ is an $N\times N$ sub-matrix of the matrix $U$ constructed by repeating the $x$th row of $U$ $n^{(k)}_x$ times, and then repeating the $y$th column of the result matrix $n^{(t)}_y$ times for all $x$ and $y$, and $\textrm{Per}(V_{k,t})$ is the permanent of the matrix $V_{k,t}$.

Finally, all $D$ probabilities in the output state $\vec{P}_{\beta} = \{{P_{\beta}}_t\} = \{|\beta_t|^2\}$ are measured by employing $N$-photon coincidence detection in $m$ modes, as described in Fig. \ref{fig:system}.

\begin{figure}[hbt!]
	\centering
	\includegraphics[width=\columnwidth]{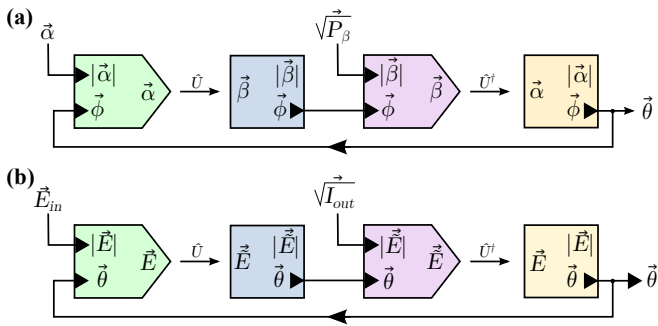}
    \caption{A schematic description of phase retrieval algorithm for (a) quantum and (b) classical light. Each iteration in the algorithm uses the input state amplitudes, $\vec{\alpha}$ or $\vec{E_{in}}$, transforms these amplitudes ($\hat{U}$), and applies the measured photon correlations $\vec{P}_\beta$ or intensities $\vec{I}_{out}$, which is followed by the inverse transformation ($\hat{U}^{\dagger}$), for the quantum or classical algorithms respectively. The object phases $\vec{\theta}$ evolve over the iterations of the algorithm, while for the quantum algorithm these phases are found from $\vec{\phi}$. \label{fig:diagram}}
\end{figure}

\subsection{Phase retrieval algorithm}

We will now describe the procedure for reconstructing the set of unknown phases $\vec{\theta}$, which generalizes the Gerchberg-Saxton (GS) error reduction iterative algorithm \cite{GerchbergAndSaxton1972} to quantum light, as shown in Fig \ref{fig:diagram}(a). The goal is to retrieve the phases $\vec{\theta}$ from the known amplitudes of the input state $\vec{\alpha}$, and the measured probabilities of the output state $\vec{P}_{\beta}$.

We begin by guessing random initial values for the set of $m$ phases, $\vec{\theta^{(0)}}$.
The $i$th iteration of the algorithm begins by constructing an input state $|\tilde{\psi}_{\vec{\theta}}^{(i)}\rangle = \sum_{k=1}^D \alpha_k \exp(\imath \phi_k^{(i-1)}) |\vec{n}^{(k)}\rangle$, as in Eq. \ref{eq:psi_theta}, where $\phi_k^{(i-1)} = \vec{\theta}^{(i-1)}\cdot\vec{n}^{(k)}$. Then, the state $|\tilde{\psi}_{\vec{\theta}}^{(i)}\rangle$ is transformed as in Eq. \ref{eq:psi_F} to find a set of output state amplitudes $\vec{\beta}^{(i)}$.  The arguments of these complex amplitudes are combined with the measured output probabilities $\vec{P}_{\beta}$ to yield a new estimate of the output state $|\psi_F^{(i)}\rangle = \sum_{t=1}^D \sqrt{{P_{\beta}}_t}\exp(\imath \, \textrm{arg}\, (\beta_t^{(i)}))|\vec{n}^{(k)}\rangle$. This state is then transformed back to retrieve the corresponding input state $|\psi_{\vec{\theta}}^{(i)}\rangle = \hat{U^{\dagger}}|\psi_{F}^{(i)}\rangle = \sum_{k=1}^D \alpha_k^{(i)} \exp(\imath \phi_k^{(i)})|\vec{n}^{(k)}\rangle$, from which a new estimate for the set of phases $\vec{\theta}^{(i)}$ is found by inverting the relation $\phi_k^{(i)} = \vec{\theta}^{(i)}\cdot\vec{n}^{(k)}$

The GS algorithm is known to always converge to a solution by means of error reduction \cite{GerchbergAndSaxton1972,Fienup:82}, not necessarily to the correct solution. In order to quantify the error with which the algorithm converges we use two different measures: the error in the Fourier output state in the $i$th iteration, $\delta P_F(i)$,
\begin{equation}\label{eq:E_F}
	\delta P_F^2(i) = \sum_{t=1}^D \left( |\beta_t^{(i)}|^2 - P_{\beta_t}\right)^2,
\end{equation}
and the phase error in the $i$th iteration, $\delta\vec{\theta}(i)$,
\begin{equation}\label{eq:E_theta}
	\delta\vec{\theta}^2(i) = \sum_{x=2}^m \mod(|\theta_x^{(i)}- \theta_x| ,2\pi)^2,
\end{equation}
where $\mod(a,b)$ is the reminder of the division $a/b$ rounded to the nearest value.

\subsection{Required conditions for uniqueness} \label{subsection:uniqueness}
In order to achieve a unique phase retrieval the input quantum state amplitudes $\vec{\alpha}$ of Eq. \ref{eq:psi_input} are chosen such that they satisfy two conditions.
\subsubsection{Avoiding trivial ambiguities}
In order to eliminate the trivial ambiguities, the input states should be chosen such that they have no symmetries of translation and reflection with respect to the phases of the modes. The simplest solution is to arrange the average photon number in these modes to break those symmetries.
\subsubsection{Phase transformation}
The object phases $\vec{\theta}$ are evaluated by first estimating the phases $\vec{\phi}$ of the quantum states that are used as the interrogating field. To uniquely determine the $m$ phases, clearly one has to start with at least $m$ basis states amplitudes $\vec{\alpha}$ that are non-zero. When $m$ initial amplitudes are used, the matrix that expresses the relations between the $m$ object phase $\vec{\theta}$ and the subset of $m$ phases that are used from $\vec{\phi}$ should have non-zero determinant, so that the phases of each mode can be uniquely extracted from the reconstructed phases of the basis vectors. In addition, since the phases of the state $\vec{\phi}$ are reconstructed up to an integer number of $2\pi$, we need to make sure that this shift remains an integer number of $2\pi$ for the object phases $\vec{\theta}$ as well.

\section{Example: Retrieval of Six Phases}

We describe here a phase retrieval problem with $m=6$ as an instructive example. We assume that the transformation $\hat{U}$ is the discrete Fourier transform (DFT) which is the most relevant one for many practical realizations.
We begin with the following quantum two-photon state ($N=2$):
\begin{align}\label{eq:psi_6}
|\psi_6\rangle=\frac{1}{\sqrt{6}}(&|2,0,0,0,0,0\rangle+ |1,1,0,0,0,0\rangle+ \nonumber\\
                                &|1,0,1,0,0,0\rangle+|1,0,0,1,0,0\rangle+ \nonumber\\
                                &|1,0,0,0,1,0\rangle+|0,1,0,0,0,1\rangle).
\end{align}
The input state of Eq. \ref{eq:psi_6} contains only six of the $D_{m=6}= 21$ configurations of two photons in six modes, all six with equal amplitudes of $1/\sqrt{6}$. This input state was constructed with care in order to fulfil the two requirements for uniqueness (section \ref{subsection:uniqueness}): First, the
state is chosen to eliminate trivial ambiguities, i.e. translation or reflection. Indeed, the input state of Eq. \ref{eq:psi_6} has non-equal average photon numbers in the various ports; the intensity ratio between the modes is $6:2:1:1:1:1$, which breaks both symmetries. Second, the input state basis is chosen such that it enables the extraction of measured object phases $\vec{\theta}$, from the phases of the quantum state $\vec{\phi}$, which are actually estimated by the algorithm.


To compare the phase retrieval performance with quantum light to that with classical light, we performed two sets of simulations, with the same set of object phases that was chosen randomly to be $\vec{\theta}_{obj}=\{0, 3.22, 4.10, 4.57, 1.35, 4.11\}$.

\begin{figure}[hbt!]
	\centering
	\includegraphics[width=\columnwidth]{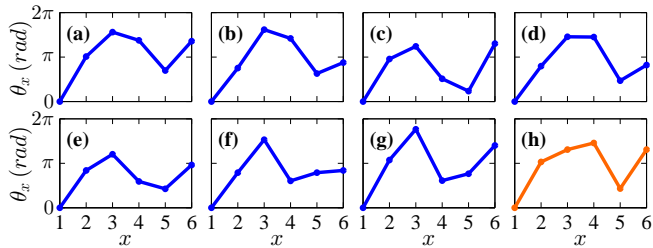}
    \caption{Different solutions found by the phase retrieval algorithm with classical light. Solutions (a)-(g) are wrong reconstructions, while (h) is the correct one. \label{fig:phase_rec_6phases}}
\end{figure}
\begin{figure}[hbt!]
	\centering
	\includegraphics[width=\columnwidth]{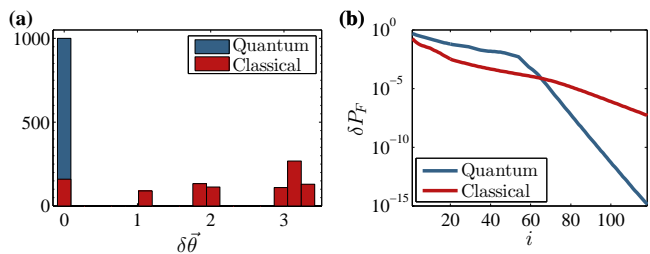}
    \caption{Comparing the performance of the phase retrieval algorithm with quantum and classical light. (a) Histograms of the retrieved phase error $\delta\vec{\theta}$ and (b) the Fourier errors $\delta P_F$ as a function of the iteration number $i$, using classical and quantum light for retrieval of $\vec{\theta}_{obj}$ for $1000$ runs of the algorithm. In the classical case, only $~16\%$ correct reconstructions were achieved, while the erroneous solutions are the majority of the instances. The Fourier error of the classical algorithm ($\delta P_F^{(cl)}$) is shown only for cases that converged to the correct solution. The quantum case, that used the entangled two-photon state given in Eq. \ref{eq:psi_6}, always converged to the correct phases. \label{fig:errors}}
\end{figure}
%

First, we applied the GS algorithm with classical light, as shown in Fig \ref{fig:diagram}(b).
We assumed a classical coherent input field with input amplitudes
$\vec{E}_{in}=\{\sqrt{6},\sqrt{2},1,1,1,1\}$, such that it reproduces the same intensity ratio as the quantum state in Eq. \ref{eq:psi_6}, and, similarly, does not have reflection and translation symmetries. This input field was transformed with the phases $\vec{\theta}_{obj}$ and then by the DFT to obtain a set of six complex output amplitudes $\vec{\tilde{E}}_{out}$. The output intensities $\vec{I}_{out}$ are then used as the input to the GS algorithm.
We ran the algorithm a large number of times, each run starting with a different random set of initial phases. The algorithm almost always converged, i.e. found a solution which reproduces the intensities in the Fourier plane with very low error in the Fourier plane, $\delta P_F\ll10^{-3}$, but most of the times it did not find the correct set of phases $\vec{\theta}_{obj}$.
All the solutions that were found using classical light are presented in Fig. \ref{fig:phase_rec_6phases}, where 7 out of 8 of these solutions are actually wrong.
A histogram showing the phase error distribution for 1000 runs of the algorithm is shown in Fig. \ref{fig:errors}(a). In this representative example, the algorithm converged to the correct phases only in about 16\% of the runs.

In contrast, the quantum algorithm always found the correct solution. The quantum simulation was performed in an analogous way: the input state of Eq. \ref{eq:psi_6} was first transformed by the phase vector $\vec{\theta}_{obj}$ according to Eq. \ref{eq:psi_theta}, and then by the DFT to calculate the set of output amplitudes $\vec{\beta}$, as in Eq. \ref{eq:psi_perm}. Note that in contrast to the $m=6$ amplitudes that characterized the classical case, here there are $D_{m=6}=21$ amplitudes, that describe all the combinations of two photons in six modes. The values of $\vec{P}_{\beta}=|\vec{\beta}|^2$ are used as input to the GS algorithm, that was run many times with random initial phases. As shown in Fig. \ref{fig:errors}(a), it always converged to the correct phase vector $\vec{\theta}_{obj}$.

Even when the algorithm using the classical light converged to the correct phases (i.e. in about 16\% of the times), it did so less efficiently than the quantum one. Fig. \ref{fig:errors}(b) shows the progressive reduction of Fourier plane error $\delta P_F$ with increasing iterations for both the quantum and classical states of light, averaged over many runs. For the quantum state, this error is given by Eq. \ref{eq:E_F}, and similarly, for the classical state it is given by $(\delta P_F^{(cl)})^2 = \sum_{x=1}^m (|\tilde{E}_x^{(i)}|^2 - |\tilde{E}_x|^2)^2 / ( \sum_{x=1}^m |\tilde{E}_x|^2)$, where $\tilde{E}_x^{(i)}$ is the far-field amplitude of the $x$th mode in the $i$th iteration \cite{Fienup:82}.

\section{Sensitivity}

In the simulations described above, the far-field amplitudes, either classical or quantum, were calculated from theory. In practice, these values would be measured. Measurements with classical or quantum states of light are often limited in sensitivity due to shot-noise. In fact, metrology with nonclassical states of light is more often than not motivated by its superior sensitivity in phase measurements. While this is not the focus of this work, for completeness, we wish to compare the precision of the algorithms with classical and quantum light. For this purpose, we performed Monte-Carlo simulations with quantum and classical light, for the same example that we discussed in the previous section, given a total number of photons passing through the sample $N_{T}$.
Again, in the classical case, we considered only the runs that yielded the correct phases which were less than $16\%$ of the total runs. In practice, of course, there is no way to identify the correct solution, but here we are interested to check the ultimate precision of the algorithm.

\begin{figure}[bth]
	\centering
     \includegraphics[width=\columnwidth]{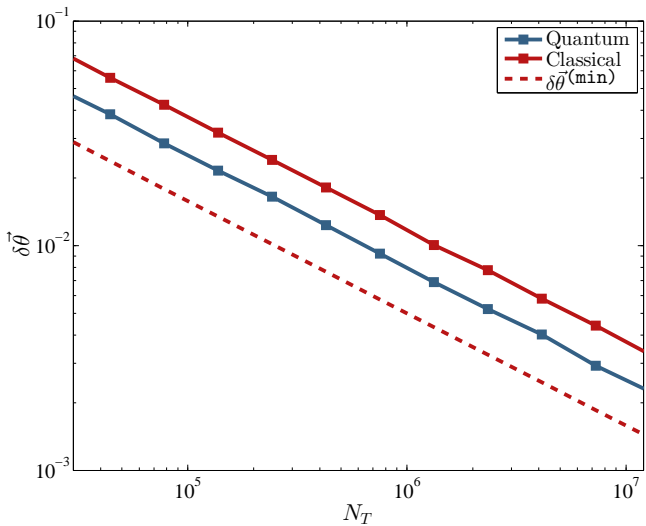}
    \caption{Phase error for the algorithm using quantum and classical light input $\delta\vec{\theta}$, and the ultimate minimal error limit achievable with classical light $\delta\vec{\theta}^{\texttt{(min)}}$, as a function of total number of photons probing the system, $N_{T}$. Here, too, The classical line is drawn only for the subset of runs, $\sim 16\% $ of all runs, that converged to the correct solution.  \label{fig:E_phi_N}}
\end{figure}
The phase error $\delta\vec{\theta}$ for the quantum and classical light are shown as a function of $N_T$ in Fig. \ref{fig:E_phi_N}. Also shown in the figure
is the minimal error that could be achieved with classical light, $\delta\vec{\theta}^{\texttt{(min)}} = (m-1)/\sqrt{N_{T}}$ \cite{MultiPhaseWalmsleyPRL2013}.
The phase error $\delta\vec{\theta}$, achieved with the particular quantum state of Eq. \ref{eq:psi_6} and for the particular phase object $\vec{\theta}_{obj}$, can be fitted by $8.0/\sqrt{N_T}$. From Fig. \ref{fig:E_phi_N} it is evident that this phase error of the quantum state is better than that achieved with classical light, $11.8/\sqrt{N_T}$, again pointing to the relative advantage of the quantum algorithm.

Our particular state does not show sensitivity below classical limits (the ultimate classical limit in our example is $5/\sqrt{N_T}$). Still it performs 48$\%$ better than classical light with the same intensity distribution, again, even if compared only with the small fraction of classical solutions that converged to the correct solution.
We note that a method for calculating the ultimate limit of sensitivity for any input pure state is given in Ref. \cite{MultiPhaseWalmsleyPRL2013}. Furthermore, enhancement factors of sensitivity greater than was shown here are probably possible, either by considering higher photon numbers $N>2$ then in our example, or by optimizing the input state amplitudes, as well as the transformation $\hat{U}$.

\section{Discussion}

The solution found with quantum light is unambiguous, converges faster, and is more precise than the one found using classical light. The reason for this lies in the fact that the quantum states are characterized by $D={{m+N-1} \choose N}$ amplitudes, which is significantly larger than the $m$ amplitudes of the classical case for any $N>1$, 21 vs. 6 in our two-photon in six modes example above. This significant increase in the number of constraints imposed on the Fourier plane quantum state amplitudes for the same number of unknown phases is most probably what leads to the elimination of extra solutions, the faster convergence, and the very small error.

We have considered in the example presented above the phase retrieval of 6 modes, however, we have checked that our approach performs perfectly also for higher number of modes. We tested our quantum algorithm with two photon states ($N=2$) for $m = 10$, $20$, and $30$ unknown phases.  For that, we have generalized the quantum state of Eq. \ref{eq:psi_6} by introducing additional terms of the form $|1,0,\dots,0,1,0,\dots,0\rangle$ having in total $m$ terms of equal amplitudes, still meeting the symmetry requirements for the input state. The algorithm performed just as well, retrieving the phases accurately for all values of $m$ tested, while the classical algorithm with coherent light with intensities that matched that of the generalized quantum state had much reduced rate of successful reconstruction, less than $1\%$ for $m\geq10$.

Finally, we discuss the considerations for practical realization of the suggested method. First, arbitrary entangled states of light are generally hard to generate, however, probabilistic generation of multi-mode correlated two-photon states is technically possible, although, to the best of our knowledge, has never been demonstrated in multi-mode systems, perhaps due to lack of interest. The DFT can be implemented by Fourier multiport devices, which have been experimentally demonstrated for quantum states of light \cite{spagnolo2012,Poem_MMW_PRL2012}. Additionally, measurements of $N$-photon correlations over $m$ modes requires a set of $D$ measurements which for large number of modes/photons can be challenging. A similar problem has been recently encountered in the realizations of the boson-sampling problem \cite{White_BS13,Walmsley_BS13,Walther_BS13,Sciarrino_BS14}. For a quantum state of two photons however, the problem involves only $m(m+1)/2$ measurements, which is quite practical, using, for example, large arrays of single photon detectors \cite{DetectingQLight_Silberhorn2007}, or cameras with single-photon sensitivities \cite{EntanglementwithCamera}.

It is important to note here that quantum measurement of correlations on classical coherent input light will not be useful: they will not yield any additional information or any other advantage, as classical states are uncorrelated and separable, unlike quantum states which can exhibit inherent photon correlations between the modes. We also note that these findings raise many interesting theoretical questions, for example on the optimal choice of the input quantum state and on its ultimate sensitivity, and how the technique will perform with two-dimensional objects, as well as objects of phase and absorption.


\section{Summary}
We studied the use of quantum states of light for phase retrieval. We showed that quantum states of two photons exhibit a few advantages in retrieving the phase of a one-dimensional object from its far-field diffracted intensity, as compared with classical states of light. The quantum approach achieves a unique reconstruction, which converges faster, and is more robust when subjected to shot-noise, when compared with classical approaches.

\section*{Funding Information}
Financial support of this research by the ERC grant QUAMI, the ICore program
of the ISF, the Israeli Nanotechnology FTA program, the Minerva foundation and the Crown Photonics Center is gratefully acknowledged. E.P. would like to acknowledge an EU Marie Curie Fellowship, a British-Technion Society Coleman-Cohen Fellowship, and the Oxford Martin
School for initial support.

\section*{Acknowledgments}

We thank Ben Leshem, Oren Raz and Dan Oron for fruitful discussions.






\end{document}